\documentclass[twocolumn,amsfonts,showpacs,superscriptaddress,nofootinbib]{revtex4-1}
\usepackage[dvipdfmx]{graphicx}
\usepackage{amssymb,amsmath,amsthm,booktabs,mathtools}
\usepackage{bbm}
\usepackage{bm}
\usepackage{color}
\usepackage{multibib}
\usepackage{tikz}
\usepackage{pgfplots}
\usepackage{enumitem}
\usepackage{hyperref}

\newcommand{\bra}[1]{{\langle #1 \vert}}

\newcommand{\ket}[1]{{\vert #1 \rangle}}

\newcommand{\ave}[1]{{\langle #1\rangle}}
\newcommand{\ii}{ {\rm i} }
\newcommand{\dd}{ {\rm d} }
\newcommand{\NN}{\mathbb{N}}
\newcommand{\ZZ}{\mathbb{Z}}

\def\tr{{\,{\rm tr}}}

\def\one{\mathbbm{1}}
\def\re{{\,{\rm Re}\,}}

\def\be{\begin{equation}}
\def\ee{\end{equation}}

\def\tr{\,{\rm tr}\,}
\def\ket#1{|#1\rangle}
\def\bra#1{\langle#1|}

\def\ave#1{\langle #1 \rangle}
\def\ii{{\rm i}}

\def\tit#1{}
\newcommand{\half}{{\textstyle\frac{1}{2}}}

\usepackage{color}

\usepackage{amsmath}	
\begin{document}

\title{The isolated Heisenberg magnet as a quantum time crystal}

\author{Marko Medenjak}
\affiliation
{Institut de Physique Th\'eorique Philippe Meyer, \'Ecole Normale Sup\'erieure, \\ PSL University, Sorbonne Universit\'es, CNRS, 75005 Paris, France}
\author{Berislav Bu\v{c}a}
\affiliation{Clarendon Laboratory, University of Oxford, Parks Road, Oxford OX1 3PU, United Kingdom}
\author{Dieter Jaksch}
\affiliation{Clarendon Laboratory, University of Oxford, Parks Road, Oxford OX1 3PU, United Kingdom}
\affiliation{Centre for Quantum Technologies, National University of Singapore, 3 Science Drive 2, Singapore 117543}

\begin{abstract}
We demonstrate analytically and numerically that the paradigmatic model of quantum magnetism, the Heisenberg XXZ spin chain, does not relax to stationarity and hence constitutes a genuine time crystal that does not rely on external driving or coupling to an environment. We trace this phenomenon to the existence of \emph{extensive dynamical symmetries} and find their frequency to be a no-where continuous (fractal) function of the anisotropy parameter of the chain. We discuss how the ensuing persistent oscillations that violate one of the most fundamental laws of physics could be observed experimentally and identify potential metrological applications.
\end{abstract}

\maketitle
Isolated systems consisting of many interacting particles are generally assumed to relax to a stationary equilibrium state whose macroscopic properties are described by the laws of statistical physics.  This has been confirmed by a large amount of recent theoretical (e.g. \cite{quench1,quench2,quench3,quench4,quench5,quench6,quench7,quench8,quench9,quench10,quench11,quench12,quench13,quench14,quench15,quench16,ETHalt}) and experimental (e.g. \cite{quenchEXP1,quenchEXP2,quenchEXP3,quenchEXP4,quenchEXP5,quenchEXP6,quenchEXP7}) work, particularly focusing on the Heisenberg spin chain in quench setups (e.g. \cite{XXZ1,XXZ2,XXZ3,XXZ4,XXZ5,EnejGGE,XXZ6,XXZ7}). On the other hand time crystals describe a phase of mater which never relaxes to stationarity and breaks the continuous time-translation symmetry (TTS) in analogy with the continuous space translation symmetry breaking in ordinary crystals. Historically the research into quantum time crystals was instigated by an intriguing possibility that a system at zero temperature could exhibit perpetual motion \cite{Wilczek}, however this has subsequently been disputed \cite{bruno1,bruno2,watanabe}, leading to possible generalisations to finite temperature, or far from thermal equilibrium. Despite the large amount of work on Floquet (breaking the discrete time translation symmetry) or dissipation induced time crystals \cite{timecrystal1,timecrystal2,Lukin,timecrystal3,timecrystal4,timecrystal5,timecrystal6,timecrystal7,timecrystal8,timecrystal9,disTC4,BucaTindallJaksch,Fazio,Esslinger,disTC2,disTC3} such behavior was believed to be impossible to realize in isolated many-body systems \cite{bruno1,bruno2,watanabe,volovik}, which typically relax to stationary states depending only on few parameters, such as energy and particle number \cite{ETHReview,VidmarRigol}. 

Despite numerous studies on relaxation in many-body quantum systems, there have been no results on spontaneous time translation symmetry breaking to date, except for mean field models (e.g. in the non-interacting limit \cite{Sascha}, fully connected models \cite{BoseHubbard}), or the case of spin precession. These systems are not believed to have the required eigenfrequency complexity for local observables to relax to stationarity \cite{ETHReview, thermoreview}. By contrast, we will focus on strongly correlated systems where one does a priori expect relaxation to stationarity \cite{quench1,quench2,quench3,quench4,quench5,quench6,quench7,quench8,quench9,quench10,quench11,quench12,quench13,quench14,quench15,quench16,ETHalt}. The absence of such results might be expected in the light of the no-go theorem \cite{watanabe}, however there is a crucial defining property underlying its derivation, which we relax.
It assumes that the system should exhibit long range spatial correlations, which are not relevant to time-translation symmetry breaking, or non-stationarity.

In this letter we show that systems can indeed fail to relax, and relate this type of behavior to \emph{extensive dynamical symmetries}, which are local in space and have a periodic dependence on time. As a consequence of dynamical symmetries systems support time-dependent statistical ensembles, and fail to equilibrate after a quantum quench. Non-stationarity shows up also on the level of dynamical response functions which describe the behavior near equilibrium, and its stability under small perturbations. 
 
We demonstrate the effects of local dynamical symmetries for the one-dimensional Heisenberg spin chain and study the stability under the integrability breaking perturbation. This paradigmatic model is used to describe many experimentally relevant situations including organic componounds \cite{exp2}, various materials \cite{exp1}, cold atom implementations \cite{coldexp}, and quantum dots \cite{dot}.

{\bf Quantum time crystals.}
Watanabe and Oshikawa defined quantum time crystals as an interacting system exhibiting persistent oscillations, which can be probed by the auto-correlation function $\tilde{f}(t)=\frac{1}{V^2} \langle O(t)O \rangle $, where $ V $ is the volume of the system and $ O $ an extensive observable \cite{watanabe}. They consider this auto-correlation function as a perturbation to equilibrium and show that it is time independent at zero and finite temperature. Their definition might not capture physically measurable persistent oscillations because the time dependent (connected) part of the function $ \tilde{f}(t) $ vanishes in the absence of long range correlations in large volume limit at all times. We instead propose a definition where the connected auto-correlation function is initially normalized $f(t)=\frac{1}{\langle O^2 \rangle} \langle O(t)O \rangle $, which probes TTS breaking if equilibrium ensemble is perturbed, even in the absence of long range correlations. Following such perturbation quantum time crystals will never reach stationarity.

While there are different ways of identifying the many-body nature of the phenomenon, the definition we use here is that single-body observables relax to stationarity, while some of the many-body observables oscillate persistently. This way the oscillation of the higher point correlation functions cannot be attributed to the single-body oscillations.

Recently, most of the work on quantum time-crystals has focused on the discrete time translation symmetry breaking in Floquet systems \cite{timecrystal1,timecrystal2,Lukin,timecrystal3,timecrystal4,timecrystal5,timecrystal6,timecrystal7,timecrystal8,timecrystal9}. In this case TTS breaking is studied in the out-of-equilibrium \emph{quench} setup. In this setup the system is prepared in a generic pure state $\ket{\psi}$ and then allowed to evolve under the action of a Hamiltonian. The system is identified as a discrete quantum time crystal if the dynamics breaks the discrete time symmetry of the driving period $T$ with a subharmonic response $\bra{\psi}o(t+nT)\ket{\psi}=\bra{\psi}o(t)\ket{\psi}$, for some integer $n>1$, and $\bra{\psi}o(t+t_1)\ket{\psi}\neq\bra{\psi}o(t)\ket{\psi}$ for $ t_1<nT $ \cite{timecrystal1}. We make the analogous identification for continuous time evolution by requiring that $\bra{\psi}o(t+T)\ket{\psi}=\bra{\psi}o(t)\ket{\psi}$ and $\bra{\psi}o(t+t_1)\ket{\psi}\neq\bra{\psi}o(t)\ket{\psi}$, for $ t_1<T $ and for some $ T $.

{\bf Extensive dynamical symmetries.}
An important insight into phenomena of equilibration is provided by the eigenstate thermalization hypothesis in generic systems \cite{ETHde,ETHSr,ETHalt,ETHReview}, or generalized eigenstate thermalization hypothesis (GETH) in integrable systems \cite{GETH}. It states that offdiagonal elements of local observables in eigenbasis of local Hamiltonian vanish exponentially in thermodynamic limit, and that their expectation values in a given eigenstate are smooth functions of conserved quantities.
Assuming the validity of GETH the system is expected to locally relax to the maximal entropy, or generalized Gibbs ensemble (GGE) $ \rho_{GGE}=\exp(-\sum_j \mu_j X_j) $, following the quantum quench. The set of chemical potentials $ \mu_j $ is obtained by matching the expectation values of extensive conservation laws $ X_j $ \cite{ChargesReview} in the ensemble and the initial state \cite{thermoreview}.
The situation is very different if the system possesses an additional set of \emph{extensive dynamical symmetries} $ Y $ and $ Y^\dagger $. Such quantities satisfy a simple closure (or eigenoperator) condition 
\begin{equation}
\label{closure_cond}
 [H,Y]=\omega Y,
\end{equation}
which leads to periodic evolution $ Y(t)=\exp(\ii \omega t)Y(0) $ and $ Y(t)^\dagger=\exp(-\ii \omega t)Y(0)^\dagger $. In any isolated system there is a large number of operators satisfying condition \eqref{closure_cond}, however in general they are highly nonlocal, and as such have no effect on local physics on large timescales. For simplicity we will discuss the case with a single frequency $ \omega $, but the generalization to multiple frequencies is straightforward.

{\bf Heisenberg model and extensive dynamical symmetries.}
Throughout the letter we will consider the anisotropic Heisenberg Hamiltonian 
\begin{align}
\begin{split}
\label{Heisenberg}
& H =J\left[(\sum_j s^x_j s^x_{j+1}+ s^y_j s^y_{j+1}+\Delta s^z_j s^z_{j+1}+\right.\\
&+\left.\alpha(\sum_j  s^x_j s^x_{j+2}+ s^y_j s^y_{j+2}+\Delta s^z_j s^z_{j+2}) \right]+\sum_j h s^z_j,
\end{split}
\end{align}
as an example of the quantum time crystal.
In equation \eqref{Heisenberg} we introduced the spin$ -\half $ operators $ s^{x,y,z} $, anisotropy $ \Delta $, hopping amplitude $ J $, the magnetic field $ h $, and the integrability breaking parameter $ \alpha $, which is set to $ 0 $ except when otherwise specified. One of the crucial aspects of the Heisenberg model, which has a paramount effect on physical properties are the extensive conservation laws \cite{Tomaz,YCharges,XCharges,ZCharges}. These conservation laws are extensive operators that commute with the Hamiltonian and have an overlap with physical local quantities. Their effects range from the absence of thermalization to the ideal energy and spin conductivity at any temperature. Despite the absence of thermalization, the Heisenberg model has in recent years served as a testbed for studying equilibration properties of strongly interacting systems \cite{XXZ1,XXZ2,XXZ3,XXZ4,XXZ5,EnejGGE,XXZ6,XXZ7}. In what follows we will show that in the easy plane regime $ -1<\Delta<1 $, it, in general, never reaches equilibrium if $ h\neq 0 $.

\begin{figure}[!]
	\begin{center}
		\vspace{0mm}
		\includegraphics[width=0.5\textwidth]{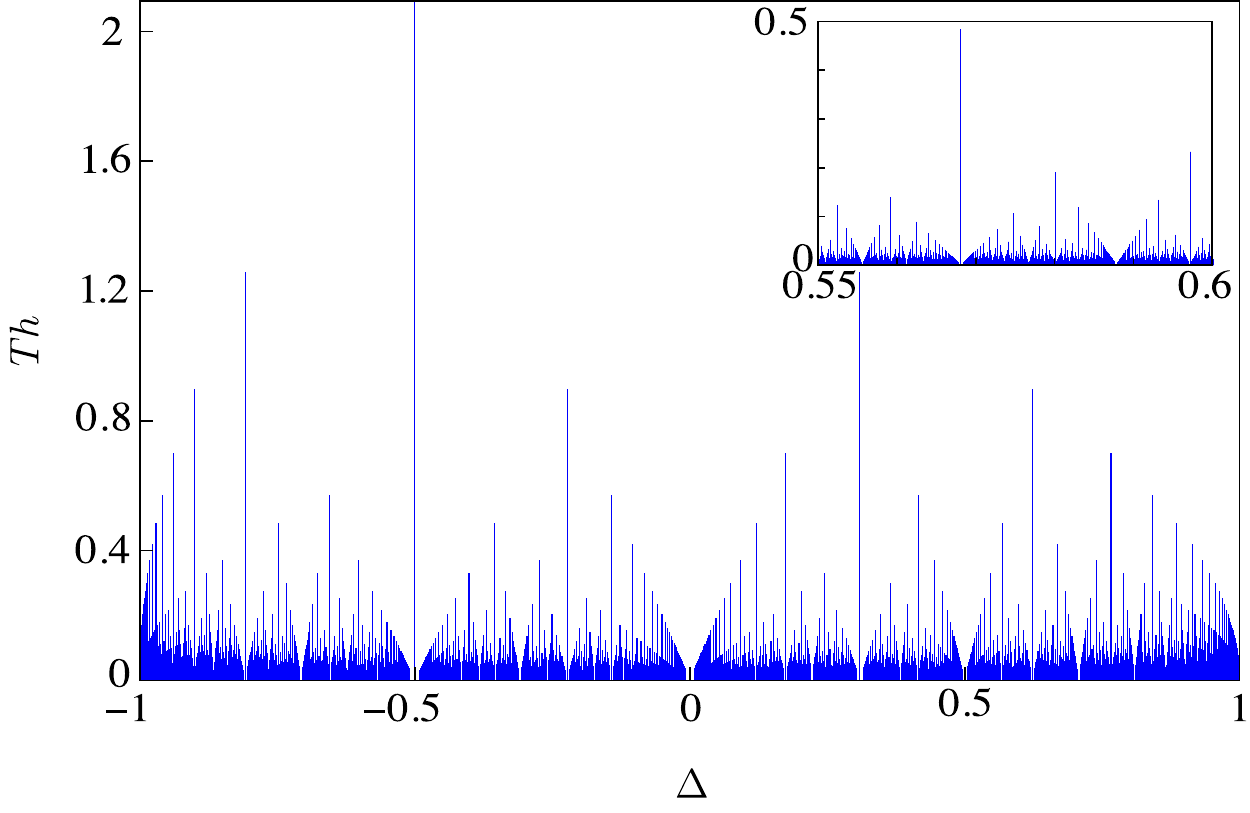}
		\vspace{-8mm}
	\end{center}
	\caption{The period $T=2 \pi/\omega$ of the persistent oscillations of the many-body observables as a function of the anisotropy $\Delta$. Inset shows a close up illustrating the no-where continuous (fractal) nature of the curve. Note the asymmetry around $\Delta=0$, which can be remedied by appropriate symmetry transformation of the quantities $Y$ \cite{YCharges}.}
	\label{fig:period}
\end{figure}

This can be seen as a consequence of semi-cyclic quantities, which were introduced in \cite{YCharges} (see also \cite{SM}).
While they commute with Hamiltonian \eqref{Heisenberg} in the absence of the field $ h=0 $, we will show \cite{SM} that they satisfy the closure condition \eqref{closure_cond} for any value of the field $ h $. Interestingly, the frequency of their oscillations $ \omega=h m $ is a discontinuous function of the anisotropy parameter $ \Delta=\cos(\frac{\pi n}{m}) $, with $ n\in 2 \NN,\ m\in 2\NN+1 $ (see FIG.~\ref{fig:period}). Moreover the explicit structure of dynamical symmetries depends finely on the exact value of anisotropy $ \Delta $, since they are comprised of densities which have a surplus of $ m $ local operators $ s^+ $. Conversely, quantities $ Y^\dagger $ have a surplus of local operators $ s^- $. This means that, for instance, at $ \Delta=-\frac{1}{2} $ we will observe the persistent oscillations of the three point transverse correlation function $ s^x_1 s^x_2 s^x_3 $, while the oscillations of this observable in the infinite time limit will be absent at any other value of $ \Delta $. Physically, such observables correspond to correlations of the three-site measurement statistics - the average measured value of each individual spin relaxes according to standard statistical physics, but the measured values will be such that on average their product oscillates in time. Alternatively, they may be thought of as oscillations of the higher moments of the $m$-site quantum fluctuations (e.g. $\ave{(s^x_1+s^x_2+s^x_3)^3(t)}$). Note that the fact that quantities responsible for oscillations Y do not exist at the non-interacting point of the model with $ \Delta=0$ solidifies the argument that the oscillations are a genuine many-body phenomena.

{\bf Time dependent generalized Gibbs ensemble (tGGE).}
In order to understand how the dynamical symmetries affect late time dynamics of local observables after the quantum quench, we consider discrete time dynamics induced by a Hamiltonian $ H $ with a period $ \frac{2 \pi}{\omega} $,
\begin{equation}
\mathcal{M}_\omega(Y_l)=\exp{(\ii\, 2 \pi H\omega^{-1})}Y_l\exp{(-\ii\, 2 \pi H\omega^{-1})}, \label{discretetimedyn}
\end{equation}
which renders $ Y $ and $ Y^\dagger $ conserved. The stationary maximum entropy ensembles for the stroboscopic dynamics can be obtained from the entropy maximization procedure, which has to respect the conservation of all conserved quantities $ X_j $, as well as dynamical symmetries $ Y $ and $ Y^\dagger $. This leads to the GGE description $ \rho_{GGE}=\exp(-\sum_j \mu_j X_j-\mu_Y Y-\bar{\mu}_Y Y^\dagger) $. If conserved quantities do not commute this might in principle require redefinition of ensembles \cite{doyon,faggoti}.
Stroboscopic time evolution of the state $ \ket{\psi(t)} $ leads to different maximum entropy states for $ t\in [0,\frac{2 \pi}{\omega}) $, which take the form of time dependent generalized Gibbs ensemble (tGGE) 
\begin{equation}
 \rho_{tGGE}=\exp(-\sum_j \mu_j X_j-\mu_Y(t) Y-\bar{\mu}_Y(t) Y^\dagger). \label{tGGE}
\end{equation} 
The values of the chemical potentials $\mu_j$, $\mu_Y(t)$ can be fixed in the following way. The maximum entropy non-stationary ensemble which correctly reproduces the initial value of $ X_k $, and the dynamics of dynamical symmetries $ Y_k $ is obtained by requiring
\begin{align}
\label{tGGE}
\begin{split}
\bra{\psi}Y_k(t) \ket{\psi}&=\tr(Y_k(t)\,\rho_{tGGE}),\\ \bra{\psi}X_k \ket{\psi}&=\tr(X_k\,\rho_{tGGE}),
\end{split}
\end{align}
An important observation is that if the following set of relations holds
\begin{equation}
\begin{aligned}
\label{consistant}
\tr(Y_k^l Y_{k'}^{l' \dagger})&=C\,\delta_{l,l'};\quad  \\ \tr(Y_k^l Y_{k'}^{l'})&=\tr(X_k^l Y_{k'}^{l'})=\tr(X_k^l Y_{k'}^{l'\dagger})=0,
\end{aligned}
\end{equation}
the constraints \eqref{tGGE} can be satisfied by the time dependent chemical potentials $ \mu_Y(t)=\mu_Y(0)\exp(-\ii \omega t) $.

{ \bf Dynamical response functions in thermal equilibrium.}
Here we focus on the response of the system in equilibrium. Extensive dynamical symmetries have profound consequences for the asymptotic behavior of dynamical susceptibilities as well. In the large time limit we expect that a local observable $ O(t) $ can be represented as a linear combination of conserved quantities and dynamical symmetries
\begin{equation}
\label{expansion1}
O(t)\underset{t\to\infty}{=}\alpha_Y^O \exp(\ii \omega t)\, Y+ h.c.+ \sum_j \alpha_j^O X_j,
\end{equation}
for the sake of calculation of the dynamical susceptibilities $ \langle O_1(t)O_2 \rangle $. This equality is valid only on the \emph{hydrodynamic} level and in the long-time limit. Here we restricted the discussion to the case of $ \langle O_1 \rangle=\langle O_2 \rangle=0 $, which can be relaxed by considering connected correlation functions. In general the coefficients $ \alpha_Y^O $ and $ \alpha_j^O $ depend, not only on observable $ O $, but also on the thermal ensemble $ \langle\bullet \rangle $. If these coefficients vanish, the observable is not expected to oscillate.

In order to obtain the dynamics of temporal correlation functions in  Heisenberg model $ \langle O_k(t) O_l \rangle $ with $ O_k=\sum_i s^x_{i} ...s^x_{i+k-1} $, we will use the ansatz \eqref{expansion1}, specializing to the infinite temperature ensemble $ \langle\bullet \rangle=\frac{\tr(\bullet)}{\tr(\one)}$. Conversely, the results provide an asymptotic solution of the quench protocol for any initial state of the form $ \rho=\sum_{k\in 2\ZZ+1} a_k\, O_k $. A particular choice of $O_k$ was made due to the non-zero overlap with the Y quantities of the Heisenberg XXZ spin chain \cite{YCharges} in the sense of \eqref{expansion1}. 

A starting point of the derivation is expression \eqref{expansion1}, where we take into account that the set of extensive conserved quantities and dynamical symmetries in the Heisenberg model is comprised of three single parameter families: unitary conservation laws $ X_s(\lambda) $, non-unitary conservation laws $ Z(\lambda) $, and semi-cyclic dynamical symmetries $ Y(\lambda) $ with the frequency which is independent of $ \lambda $ \cite{ChargesReview}. In canonical ensemble the observable $ O_k $ can be described asymptotically as
\begin{equation}
\label{exp}
O_k(t)= \int \dd \lambda\,(\exp(\ii \omega t) f(\lambda) Y(\lambda)+\exp(-\ii \omega t) \bar{f}(\lambda)Y^\dagger(\lambda)).
\end{equation}
Note that the judicious choice of the $O_k$ and the canonical ensemble means that the charges $X_k(\lambda)$ and $ Z(\lambda) $ are not relevant via \eqref{expansion1}.
The function $ f(\lambda) $ and its complex conjugate $ \bar{f}(\lambda) $ can be obtained by projecting the expression \eqref{exp} onto the set of dynamical symmetries
\begin{equation}
\label{syseq}
\langle O_k Y^\dagger(\lambda') \rangle=\int \dd \lambda f(\lambda) \langle Y(\lambda)Y^\dagger(\lambda')\rangle
\end{equation}
Specializing to the infinite temperature, the overlaps $\langle O_k Y^\dagger(\lambda')\rangle $, and kernels $\langle Y(\lambda)Y^\dagger(\lambda')\rangle$ were obtained in \cite{YCharges}. The system of equations \eqref{syseq} can be reduced to the convolution equation and solved in Fourier space (see \cite{SM} for more details). We also note that this choice of temperature implies that $\ave{O^2} \propto V$. Using the solution $ f(\lambda) $, we can calculate the susceptibility matrix
$
\langle O_k(t)O_l \rangle\underset{t\to\infty}{=}C \cos(\omega t)$,
with
$C=\left| 2 \int \dd \lambda f(\lambda) \langle Y(\lambda)O_l \rangle \right|$.
For instance, setting the anisotropy parameter to $ \Delta=-\frac{1}{2} $, and $ k=l=3 $ we get $C= \frac{1}{64} \left(\frac{27 \sqrt{3}}{\pi }-8\right) $. In general the result is expressed as a double integral of elementary functions (see \cite{SM}), and the constant which can be calculated efficiently.

\begin{figure}[!]
	\begin{center}
		\vspace{0mm}
		\includegraphics[width=0.5\textwidth]{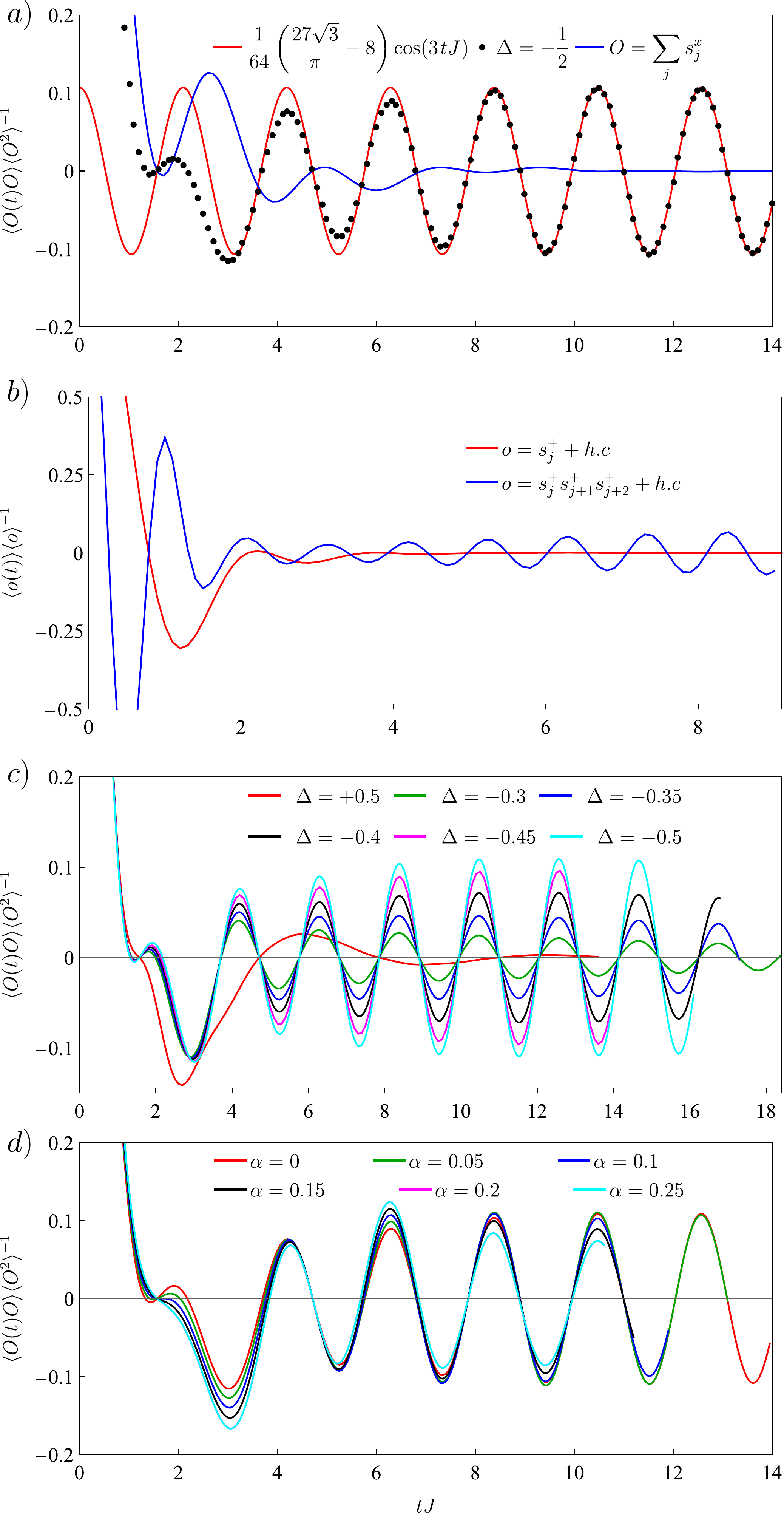}
		\vspace{-8mm}
	\end{center}
	\caption{In figure (a) we plot the non-stationary behavior of the dynamical susceptibility for $O=\sum_j s^x_js^x_{j+1}s^x_{j+2}$ for $ \Delta=-1/2,\ h=J $ and compare it to the exact result. As expected, single body observable $ O=\sum_j s^x_j $ relaxes to stationarity. In figure (b) we plot the data for the quench from the feromagnetic state, showing non-stationarity of three point function $ o=s_j^+ s_{j+1}^+ s_{j+2}^++s_j^- s_{j+1}^- s_{j+2}^- $ and relaxation of a single point function $o=s_j^++s_j^- $, for $ \Delta=-0.5,\ h=2 J $. In figure (c) we present the effects of perturbing $ \Delta $ at $h=J $ on the dynamical susceptibility for $O=\sum_j s^x_js^x_{j+1}s^x_{j+2}$. In figure (d) we present the effects that integrability breaking next-to-nearest neighbor interaction $ \alpha $ has on the dynamical susceptibility for $O=\sum_j s^x_js^x_{j+1}s^x_{j+2}$ at $ \Delta=-1/2,\ h=J $. All simulations were performed using DMRG with the system size $N=100$.}
	\label{fig:autocor}
\end{figure}

{\bf Numerical results.}
In Fig.~\ref{fig:autocor} a) we plot the time evolution of the autocorrelation function $C(t)=\frac{\ave{O(t)O}}{\ave{O^2}}$, for the observable $ O=\sum_j s_j^x s_{j+1}^x s_{j+2}^x $, and the infinite temperature ensemble $ \langle\bullet \rangle=\frac{\tr(\bullet)}{\tr(\one)} $. Clearly $ C(t) $ does not equilibrate, and we can compare the numerical result with the analytical calculation $C(t)\underset{t\to\infty}{=}\frac{1}{64} \left(\frac{27 \sqrt{3}}{\pi }-8\right) \cos(3 h t)$, obtaining a perfect agreement. As predicted by theory, we observe no oscilations of the transverse magnetization $ O=\sum_j s_j^x $.

In Fig.~\ref{fig:autocor} b) we plot the time dependence of a local three point correlation function  $ o=s_j^+ s_{j+1}^+ s_{j+2}^++s_j^- s_{j+1}^- s_{j+2}^- $ from the ferromagnetic initial state maximally polarized in the $ x $ direction,
\begin{equation}
\label{init_state}
\ket{\psi}=2^{-N/2}(\ket{\uparrow}+\ket{\downarrow})^{\otimes n},
\end{equation}
as well as the one-point function $o=s_j^++s_j^- $ (single body observable). It shows the relaxation of the one-point function and persistent oscillations in the three point function, illustrating the many body nature of the time crystalline behaviour.

As shown in Fig.~\ref{fig:autocor} c) we find numerically that the oscillations of relevant observables persist for a long time with an altered amplitude, following  small to intermediate perturbations of the anisotropy. At $\Delta\neq-0.5$ the three-point operator $s^x_j s^x_{j+1}s^x_{j+2}$ no longer has an overlap with the $Y$ ($\tr( Y s^x_j s^x_{j+1}s^x_{j+2})=0$, ) and thus does not pertain to any dynamical symmetry (see Eqs.~\eqref{tGGE},\eqref{expansion1}). Only at a relatively large value of perturbation $ \delta \Delta = 0.2 $ do we see significant damping of the amplitude on the accessible time scales. The oscillations are stable also with respect to the integrability breaking term $ \alpha $ on accessible timescales as shown in Fig.~\ref{fig:autocor} d). We use the general results of \cite{Eckstein,Mori} to find that the $Y$ operators are conserved under stroboscopic time evolution in Eq. \eqref{discretetimedyn} up to at-least second-order in perturbation strength (see also \cite{stability}). This implies that the oscillations decay no faster than $\exp(-tJ  {\cal O}(\alpha^2))$. Though this may possibly be improved as in the cases where the bounds on prethermalization are exponential \cite{Abanin,Mori,KAMPRX}.

{\bf Experimental realization.}  Due to the demonstrated stability, we expect that in current quantum cold-atom simulations of the XXZ spin chain, such as those done in I. Bloch's group \cite{simul1}, the lattice depth can be sufficiently tuned to make the dynamics fast enough compared to integrablity-breaking effects to observe oscillations. 

Measurement of local on-site equal-time many-body correlation functions, such as the ones we study, is available through quantum gas microscopes for cold atom systems \cite{coldexp}. For experiments an important discovery is that oscillations can be observed for a quench from the ferromagnetic initial state \eqref{init_state}, which can be engineered \cite{coldexp}. Preparation and measurement of auto-correlation functions is more involved, but can be achieved through the use of ancilla qubits in Rydberg atoms (see \cite{SM} for details). Our results could potentially also have far-reaching applications in quantum metrology \cite{metro}, as they offer, in theory, \emph{infinitely}
 sensitive and precise measurement of the system anisotropy. In cold atom simulations this can be directly related to the strength of the external magnetic field used to achieve Feshbach resonance of the spin-spin interaction \cite{coldexp}.
In this regard, an important observation is that the amplitude seems to be less effected by integrability breaking, then by the change of anisotropy.


{\bf Conclusion:}
Numerous questions remain open. While we have addressed the question of stability to perturbations from a practical perspective, stability to all orders remains an open problem, related to the long standing question of the existence of KAM theorem in systems with infinitely many degrees of freedom \cite{kappeler2013kdv}. That being said, the crucial ingredient for oscillations is not integrability itself, but rather local or extensive quantities satisfying the relation \eqref{closure_cond}. Importantly, we were able to identify similar quantities in topological models \cite{topological}, however we postpone the in-depth discussion on this topic to later publications. The glimmers of similar dynamical symmetries have also been identified in locally constrained models exhibiting quantum many-body scars, preventing the systems from relaxing for certain initial conditions \cite{scars,scars1,scarsexp}.

Some questions remain also from the standpoint of integrable systems. Here we only focused on the lowest frequency of oscillations at a given $ \Delta $, while in general the state $ \rho_{tGGE} $ should support a complete harmonic spectrum $\omega= k h m $, for $ k \in \NN $. Furthermore, due to the noncommutativity of the conserved quantities and dynamical symmetries, subtleties might arise in obtaining the correct form of $ \rho_{tGGE} $ \cite{doyon}. The answers to these questions should be attainable by extending thermodynamic Bethe ansatz description  \cite{string_charge}, to include additional quantities $ Y $. The existence of these quantities implies that the standard GGE description is in fact incomplete even in the absence of persistent oscillations ($h=0$). Another exciting question is whether the dynamical symmetries, and absence of many-body equilibration has a counterpart in the realm of classical physics. Otherwise, the phenomenon would constitute one of the first many-body quantum effects that can be observed in macroscopic systems on large space and time scales solely due to the extensive many-body nature of the $ Y $ operators.

\begin{acknowledgments}
We thank D. Bernard, J. De Nardis, E. Ilievski, T. Prosen, and L. Zadnik for comments on the manuscript.
MM is grateful to T. Prosen for granting him access to computational resources, and thanks J. De Nardis for stimulating discussions.  DMRG calculations were performed using the ITensor Library \cite{Itensor}. BB and DJ acknowledge funding from EPSRC programme grant EP/P009565/1, EPSRC National Quantum Technology Hub in Networked Quantum   Information Technology (EP/M013243/1), and the European Research Council under the European Union's Seventh Framework Programme (FP7/2007-2013)/ERC Grant Agreement no. 319286, Q-MAC. 
\end{acknowledgments}

\bibliographystyle{apsrev4-1}
\bibliography{TC}


\onecolumngrid
\newpage

\renewcommand\thesection{S\arabic{section}}
\renewcommand\theequation{S\arabic{equation}}
\renewcommand\thefigure{S\arabic{figure}}
\setcounter{equation}{0}

\begin{center}
	{\Large \emph{Supplementary information}: The isolated Heisenberg magnet as a quantum time crystal}
\end{center}

\renewcommand\thesection{S\arabic{section}}
\renewcommand\theequation{S\arabic{equation}}
\renewcommand\thefigure{S\arabic{figure}}
\setcounter{equation}{0}

\section{Semi-cyclic dynamical symmetries}
A starting point for constructing the conservation laws in Heisenberg model is the Lax matrix \cite{ChargesReview_ap}
\begin{equation}
L(\lambda)=\frac{1}{\sinh\eta}   \begin{bmatrix}
\sin(\lambda+\eta\, {\bf S}^z) & \, \sin(\eta){\bf S}^- \\
\sin(\eta) {\bf S}^+ & \sin(\lambda-\eta {\bf S}^z)
\end{bmatrix}.
\end{equation}
where the operators acting on the auxiliary space $  {\bf S}^\alpha$ have to satisfy the $ U_q(\mathfrak{sl}_2) $ quantum group relations \cite{QGBook_ap}
\begin{align}
\label{alg}
[{\bf S}^+,{\bf S}^-]&=[2 {\bf S}^z]_q, \\
q^{2{\bf S}^z}{\bf S}^{\pm}&=q^{\pm 2}{\bf S}^\pm q^{2 {\bf S}^z}.
\end{align}
The constant $ \eta $ relates to the anisotropy as $ \Delta=\cos(\eta) $.
There are three distinct families of extensive conserved quantities in the anizotropic Heisenberg model if the magnetic field is absent $ h=0 $. For our purpose, the relevant family is the family of semicyclical conservation laws \cite{YCharges_ap}, which exist for the value of $ \eta=\frac{\pi n}{m} $, and cover the interval $ -1<\Delta<1$ densely. These conservation laws are obtained from the (semi)cyclic representation of $ U_q(\mathfrak{sl}_2) $ \cite{korf}
\begin{align}
\label{qdefsc}
&{\bf S}^z=\sum_{n=0}^{m-1}(s-n) |n\rangle \langle n|,\\
&{\bf S}^+=\sum_{n=0}^{m-2}\left([k+1]_q+\frac{\alpha \beta}{[2s-k]_q}\right)|n+1\rangle \langle n|+\alpha |m-1\rangle\langle 0|,\\
&{\bf S}^-=\sum_{n=0}^{m-2}[2s-k]_q |n\rangle \langle n+1|+\beta |0\rangle \langle m-1|,
\end{align}
where the $ q $-deformation of a number $ [a]_q $ corresponds to
$
[a]_q=\frac{q^a-q^{-a}}{q-q^{-1}}$. Note that quantities are conserved only for even $ n $ and odd $ m $. Conserved quantities are extensive in the region $ \re(\lambda)\in \{\frac{\pi}{2}-\frac{\pi}{m},\frac{\pi}{2}+\frac{\pi}{m}\}$,  if the representation parameters $ s=0 $, and the derivative of the transfer matrix $ T(\lambda) $ is taken with respect to parameter $ \beta $ at $ \beta=\alpha=0 $ \cite{YCharges_ap}
\begin{equation}
Y(\lambda)=[\partial_{\beta} T(\lambda)]_{\beta=0};\quad T(\lambda)=\tr_a(L(\lambda)_{a,1}\cdot L_{a,2}(\lambda)\cdot... \cdot L_{a,n}(\lambda)).
\end{equation}
What is important for our purpose is that this set of conservation laws does not commute with magnetization and as a consequence with Heisenberg Hamiltonian if $ h\neq 0 $. It is, however, rather simple to show that they satisfy the closure condition from the main text (1). This is a consequence of operators $ Y(\lambda) $ having a surplus of operators $ s^+ $. The local densities of magnetization are supported on a single site, implying that we can calculate the commutator by considering a single site commutators of $ s^z $ with $ s^{\pm,z,0} $, $ [s^z,s^{\pm,z,0}]=\alpha^{\pm,z,0} s^{\pm,z,0} $. Since $ \alpha^{z,0}=0 $, and $ \alpha^{\pm}=(-1)^{\frac{1\mp 1}{2}} $ and $ Y $ comprises operators with a fixed surplus $ m $ of $ s^+ $ we have
\begin{equation}
[\sum_i s^z_i,Y(\lambda)]=m Y(\lambda).
\end{equation}

Let us now briefly discuss some additional properties of quantities $ Y(\lambda) $. First of all, it is obvious that $ Y(\lambda)Y^{\dagger}(\lambda) $ is conserved.
Interestingly enough, for finite sizes quantities $ Y(\lambda) $ do not commute with $ Y^\dagger(\lambda) $. This property is interesting, since the commutator of two local operator is a local operator, and we might expect that the commutator of two quasilocal operators is quasilocal as well. This, however, is nontrivial to show. Furthermore, the commutator of the operators $ A(\lambda,\mu)=[Y(\lambda),Y^\dagger(\mu)] $, is spin flip $ S=\sigma^x_1\otimes \sigma^x_2\otimes...\otimes \sigma^x_n $ antisymmetric $ S A(\lambda,\mu)S=-A(\lambda,\mu) $ by construction, and conserved for any value of the field $ h $. Provided that the commutator is extensive it should be expressible in terms of charges $ Z(\lambda) $ \cite{Prosen_periodic}, $ A(\lambda,\mu)=\int \dd \nu\, h(\nu) Z(\nu) $, since the set of the charges $ Z(\lambda) $ seems to be complete \cite{Zchargi_app,Prosen_periodic,EnejJacopo,Delucacharges,LZP}. Furthermore, one can check that quantities $ Y(\lambda) $ do not commute with $ A(\lambda,\mu) $ for small systems, and could in principle lead to new $ U(1) $ symmetry breaking charges. We, however, do not think that this is the case for two reasons: firstly numerically obtained dynamical susceptibility describes the dynamical correlation functions perfectly for $ \Delta=-\frac{1}{2} $, and secondly semicyclic charges are closely related to the current carrying charges $ Z(\lambda) $, for which, as already mentioned, there is a lot of evidence indicating their completeness. 

\section{Time dependent dynamical susceptibilities}
In this section we discuss how to analyticaly obtain the asymptotic values of time dependent susceptibilities (autocorrelation functions) for observables $ O_k=\sum_i \sigma^x_i...\sigma^x_{i+k-1} $, using the method of hydrodynamical projection \cite{hydro}. First we focus on the infinite temperature case, and later discuss the general GGE case. The starting point is the expression (8) from the main text, which after the multiplication with $ O_l $ from the right yields
\begin{equation}
\langle O_k(t) O_l \rangle=C_{k,l,m} \cos(\omega t),
\end{equation}
up to the phase factor, and $ m $ corresponds to the parametrization of anisotropy $\Delta =\cos(\tfrac{\pi n}{m})  $. The constant $ C_{k,l,m} $ is
\begin{equation}
C_{k,l,m}=\left| 2 \int \dd \lambda f_{k,m}(\lambda) \langle Y(\lambda)O_l \rangle \right|.
\end{equation}
Note that we reintroduced indices $ k $ and $ m $ in the function $ f $, and that $ Y(\lambda) $ are $ m $ dependent as well.

Let's now focus on solving the equation (9) from the main text, and in turn provide an explicit expression for constants $ C_{k,l,m} $. The starting point is the kernel derived in \cite{YCharges_ap}
\begin{equation}
\kappa_m(\lambda,\mu)=\langle Y(\lambda) Y^\dagger(\mu) \rangle=\frac{\sin\lambda\, \sin\bar{\mu}}{2\sin^2 \eta} \frac{\sin(\lambda+\bar{\mu})}{\sin(m(\lambda+\bar{\mu}))}.
\end{equation}
Using the conjecture that the complete set of charges can be obtained as a power expansion of the continuous family $ Y(\lambda) $ along the line $ \lambda=\frac{\pi}{2}+\ii\, t $ wrt. $ t $, we can restrict the discussion to the family of quantities $ Y(\lambda) $ with a single real parameter
\begin{equation}
K_m(\lambda,\mu)=\kappa_m(\tfrac{\pi}{2}+\ii \lambda,\tfrac{\pi}{2}+\ii \mu)=\frac{\cosh(\lambda) \cosh(\mu)}{2\sin^2 \eta} \frac{\sinh(\lambda-\mu)}{\sinh(m(\lambda-\mu))}.
\end{equation}
The overlap of $ Y(\lambda) $ with observable $ O_k $ reads \cite{YCharges_ap}
\begin{equation}
O_{k,m}(\lambda)=\langle O_k Y^\dagger \rangle=C(m,k)(\cosh \lambda)^{-k+2}.
\end{equation}
This leads to a set of Fredholm equation
\begin{equation}
\int \dd \lambda\, \frac{\cosh(\lambda)\cosh(\mu)}{2 \sin^2\eta}\frac{\sinh(\lambda-\mu)}{\sinh(m(\lambda-\mu))}f_{k,m}(\lambda)=C(m,k)(\cosh \mu)^{-k+2},
\end{equation}
which can be reduced to the set of convolution equations by rescaling the function $ f_{k,m}(\lambda)=C(m,k)\frac{2 \sin^2\eta}{\cosh(\lambda)}f_{k,m}'(\lambda) $:
\begin{equation}
\label{conveqn}
\int \dd \lambda\, \frac{\sinh(\lambda-\mu)}{\sinh(m(\lambda-\mu))}f_{k,m}'(\lambda)=(\cosh \mu)^{-k+1},
\end{equation}
and solved in Fourier space
\begin{equation}
\tilde{f}_{k,m}'(\zeta)=\frac{1}{\sqrt{2 \pi}}\frac{\tilde{A}_{k}(\zeta)}{\tilde{K}_{m}(\zeta)}.
\end{equation}
Here we introduced the Fourier transform of the rhs \eqref{conveqn} $ \tilde{A}_k(\zeta)=FT[(\cosh \lambda)^{-k+1}] $, and conjecture its analytical form 
\begin{equation}
\tilde{A}_{k+1}(\zeta)=
\begin{cases}
\Pi_{l=1}^{k/2-1}(\zeta^2+(2l)^2)\zeta\frac{\sqrt{\frac{\pi }{2}} \text{csch}\left(\frac{\pi  \zeta }{2}\right)}{(k-1)!}; \quad & k\in 2\ZZ  \\
\Pi_{l=1}^{(k-1)/2}(\zeta^2+(2l-1)^2)\frac{\sqrt{\frac{\pi }{2}} \text{sech}\left(\frac{\pi  \zeta }{2}\right)}{(k-1)!}; \quad & k\in 2 \ZZ+1
\end{cases} 
\end{equation}
The kernel in Fourier space reads $ \tilde{K}_m(\zeta)=FT[\frac{\sinh(\lambda)}{\sinh(m\lambda)}] $,
\begin{equation}
\tilde{K}_m(\zeta)=\frac{\sqrt{\frac{\pi }{2}} \tan \left(\frac{\pi }{m}\right)}{m \left( \frac{\cosh \left(\frac{\pi  \zeta }{m}\right)}{\cos\left(\frac{\pi}{m} \right)}+1\right)},
\end{equation}
This implies that we can write the dynamical structure constants as
\begin{equation}
C_{k,l,m}=\left|4\,C(m,l)\,\sin^2\eta \int \frac{\dd\lambda}{\cosh(\lambda)} \int \dd\zeta \exp(\ii \zeta \lambda)\tilde{f}_{k,m}'(\zeta) \int \dd \zeta' \exp(-\ii \zeta' \lambda) \tilde{A}_{l-1}(\zeta') \right|.
\end{equation}
and finally the integration over $ \dd \lambda $ yields
\begin{equation}
C_{k,l,m}=\left|4 \pi\,C(m,l)\,\sin^2\eta\, \int \dd \zeta\, \dd \zeta' \frac{\tilde{f}_{k,m}'(\zeta) \tilde{A}_{l-1}(\zeta')}{\cosh(\frac{\pi}{2}(\zeta-\zeta'))} \right|.
\end{equation}

Specializing to the value $ \eta=\frac{2\pi}{3} $, we conjecture the value of the constant in the expression for $ O_{k,m}(\lambda) $ is $ C(3,k)=\sqrt{3}\, 3^{k-3}/16^{2^{(k-m)}}$. In particular, if we set $k=l=3  $, we get the final result $ C_{3,3,3}=\frac{1}{64} \left(\frac{27 \sqrt{3}}{\pi }-8\right) $.
The function $ C(k,m) $ can be obtained by evaluating $ k $-th power $ C(k,m)=(-1)^{(k-m)}\bra{m-1}\mathbf{X}(\eta)^k \ket{1}$ of matrix
\begin{equation}
\mathbf{X}(\eta)=\half \left(\sum_{k=0}^{m-2} \sin(k \eta)\ket{k}\bra{k+1}+\sin((k+1) \eta)\ket{k+1}\bra{k} \right).
\end{equation}
In general we can show for $O$ with the smallest support with non-zero constant (namely, $k=m$) that $C(m,m)=\frac{1}{2^m} \prod_{\mu=2}^{m-1} \sin(\mu \eta)$.

\begin{figure}[!]
	\begin{center}
		\vspace{0mm}
		\includegraphics[width=0.5\textwidth]{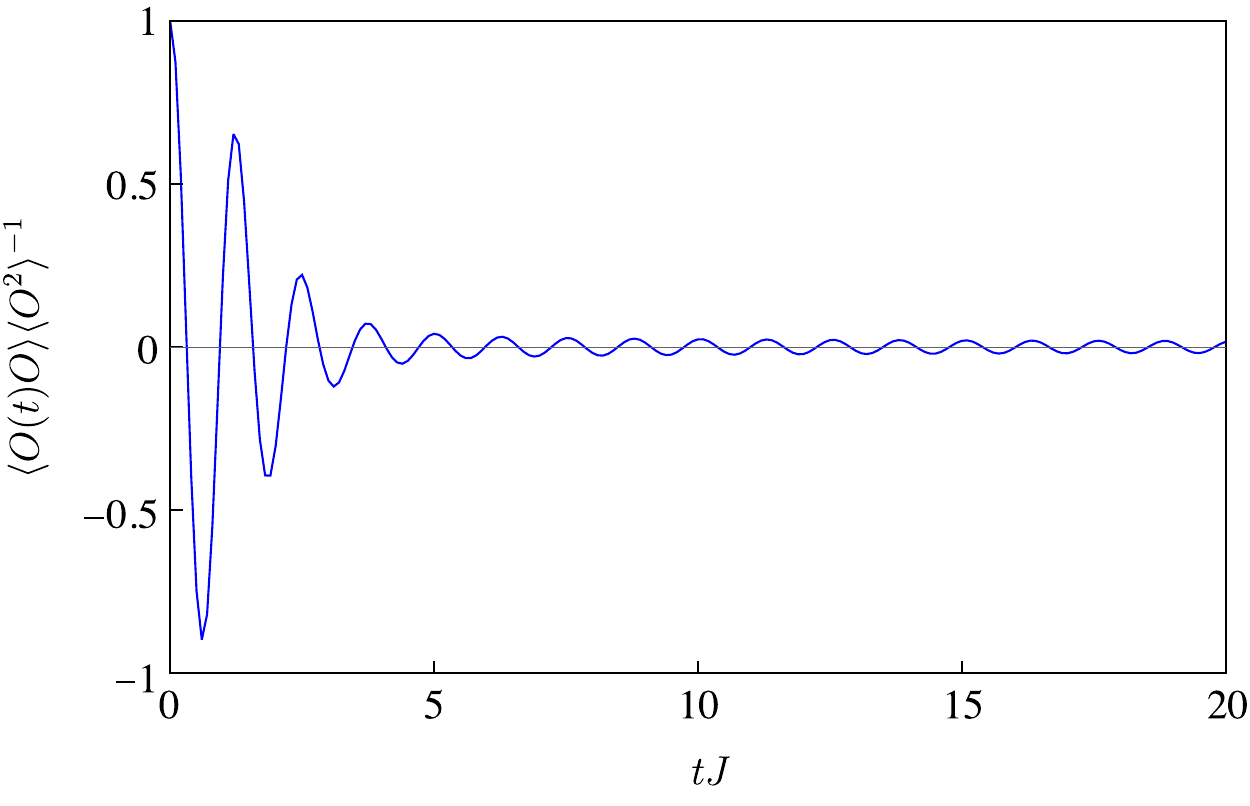}
		\vspace{-8mm}
	\end{center}
	\caption{The autocorrelation function for $O=\sum_j s^x_js^x_{j+1}s^x_{j+2}s^x_{j+3}s^x_{j+4} $ at $\Delta=\cos\left(\tfrac{2 \pi}{5}\right)$ at infinite temperature. The period of the persistent oscillations in consistent with the predicted $\omega=5J$. 
	}
	\label{figsup}
\end{figure}

To illustrate another value of $
\Delta$ in Fig.~\ref{figsup} we show the autocorrelation function for $O=\sum_j s^x_js^x_{j+1}s^x_{j+2}s^x_{j+3}s^x_{j+4}$ at $\Delta=\cos\left(\tfrac{2 \pi}{5}\right)$. 

\section{Measurement of correlation functions}

In this discussion we focus on cold atom experimental setups \cite{coldexp_ap}, though other implementations may be possible. The measurement of the local time correlation functions, such as $o=\ave{s^+_is^+_{i+1}s^+_{i+2}}+h.c$ from the main text, can be achieved by on-site quantum gas microscopy, as is discussed for two-point correlation functions in \cite{coldexp_cor} in the Fermi-Hubbard model. Combining existing techniques for simulating the XXZ spin chain with cold atoms \cite{coldexp_ap} with quantum gas microscopy would be required.  Measurement of the examples for $o$ from the main text can be done by collecting the statistics of repeated measurements from the same initial state by measuring the local Hermitian observables making up $o$, e.g. $s^x_i s^y_i s^x_i$, etc. 

Measurement of the autocorrelation functions of the form $\ave{O(t)O}/{\ave{O}^2}$ is more involved, but it could be achieved in three ways utilizing current technologies. 

Firstly, one may collect the statistics of the joint probability distribution for $O$ at different times. Preparing the initial infinite temperature state could be accurately achieved by preparing a very high initial temperature state. This should be followed by an application of the external magnetic field in the z-direction to avoid introducing $s^z_i$ terms in the initial density matrix. 

Secondly, we may also easily observe that measuring $\ave{O(t)O}$ is equivalent to measuring $\ave{O(t)}$ for a quench from an initial density matrix of the form $\rho(0)=\exp{\left( \mu O \right)}$ for very small $\mu$ (linear response regime). Such an initial state could be prepared with Rydberg atoms. Introducing ancilla qubits to the XXZ spin chain setup would mimic local three-site interactions of the form $O=\sum_is^x_is^x_{i+1}s^x_{i+2}$, as was proposed for a transverse-field Ising model, but in 2D \cite{Alex}. By tuning the external field as discussed in \cite{Alex} we can make the $a/\mu$ in $H=a H_{\rm XXZ}+\mu O$ very small for small $\mu$, thus realizing a state close to $\rho(0)$. The field would then be quenched \cite{Alex} such that $a$ becomes large and a homogenous magnetic field in the z-direction introduced. One may then measure $O(t)$ with on-site techniques, as discussed previously. 

Thirdly, one can use the method of single-qubit interferometry \cite{inter} to measure the autocorrelation by measuring a probe qubit as discussed in \cite{Juha} for measuring the Green's function in a strongly-correlated electronic system. This would require adapting the same protocol for spins.


\end{document}